\documentclass[twoside,english,final,5p,times,twocolumn]{elsarticle}
\usepackage{lineno,hyperref}
\usepackage[T1]{fontenc}
\usepackage[latin9]{inputenc}
\usepackage{amsmath}
\usepackage{amssymb}
\usepackage{graphicx}

\makeatletter
\journal{Physics Letters B}

\makeatother

\usepackage{babel}
\begin{document}
\begin{frontmatter}
\title{Particle number projection on a spatial domain}
\author[lmr,lmr1]{L.M. Robledo\corref{cor1}}
\ead{luis.robledo@uam.es}
\cortext[cor1]{Corresponding author}
\address[lmr]{Dep Física Teórica and CIAFF, Facultad de Ciencias, Universidad Autónoma
de Madrid, E-28049 Madrid, Spain}
\address[lmr1]{Center for Computational Simulation, Universidad Politécnica de Madrid,
Campus de Montegancedo, Boadilla del Monte, E-28660-Madrid, Spain}
\begin{abstract}
The formalism of particle number on a spatial domain for mean field
wave functions with pairing is revisited to account for the case where
finite dimensional basis are used. The formulas differ from the ones
previously used in the literature. It is shown that the present formalism
has the right limit in the well known case of zero pairing whereas
the other formalism do not satisfy this basic requirement. By using
a simple one-dimensional model we illustrate the differences in the
results for particle number distribution probability obtained with
the two methods. 
\end{abstract}
\begin{keyword}
Fission, fragment properties, symmetry restoration
\end{keyword}
\end{frontmatter}

\section{Introduction}

Both in nuclear fission and in reaction dynamics one has to deal with
situations in the different stages of dynamical evolution where the
system is composed of two or more fragments each of them to be treated
as individual systems. The properties of those fragments have to be
extracted from the total wave function of the system in a quantum
mechanical consistent way. For instance, the determination of the
(integer) number of particles in each of the fragments represents
a problem: the total number of particles of the system is obviously
an integer number but the probabilistic character of quantum mechanics
and its characteristic probability density distribution does not guarantee
that the same property holds for the fragments. The problem is aggravated
if pairing correlations are present in the system and the optimal
mean field treatment departs from the traditional Hartree-Fock (HF)
theory and the Hartree-Fock-Bogoliubov (HFB) method with its characteristic
breaking of the particle number symmetry is considered. The problem
how to restore particle number symmetry in a fragment was first addressed
by Simenel in Ref \cite{PhysRevLett.105.192701} were the use of
a projector on particle number in a given region of space was introduced.
The formalism was applied to the description of fusion reactions with
Slater determinants describing the two fusioning nuclei. Later, the
formalism was extended in Ref \cite{Scamps2013} to include Bardeen-Cooper-Schieffer
(BCS) wave functions and the evaluation of overlaps was carried out
with the Pfaffian formalism \cite{Robledo2009,Robledo2011,Bertsch2012}
to keep track of the sign of the overlaps. In the next years a flurry
of publications dealing with the problem has appeared \cite{Scamps2013,PhysRevC.100.024612,PhysRevC.100.034612,PhysRevC.101.014310,PhysRevC.103.054602,PhysRevC.104.054601,PhysRevLett.126.142502,Schunck2022,Bulgac2022,Scamps2017,Marevic2021,Qiang2025}
and the formalism extended even to deal with the restoration of rotational
symmetry in the fragment \cite{PhysRevC.104.054601,Marevic2021}.
These developments have lead to another avalanche of papers discussing
various effects related to fission fragment's properties due to particle
number restoration. In these publications, the formalism developed
for HFB or BCS wave functions runs parallel to the old formalism developed
to restore particle number symmetry \cite{Bayman1960,Dietrich1964,Mang1966,Ring2000}
with the only modification of replacing the particle number operator
$\hat{N}$ in the gauge angle rotation operator $e^{i\varphi\hat{N}}$
by its counterpart $\hat{N}_{\mathcal{D}}$ restricted to a very specific
region of space (called a spatial domain $\mathcal{D}$ ). Unfortunately,
those papers usign BCS or HFB overlook the impact of using finite
basis in the calculations \cite{Robledo1994,Rob22} and therefore
they may obtain incorrect expressions for the required quantities as it
will be shown below. 
It turns out that the expression used in \cite{Scamps2013} 
using the results of \cite{Bertsch2012} based on pfaffians is applicable to the 
present case even thought the formalism of \cite{Bertsch2012} assumes implicitly
a basis closed under the symmetry operation. The reason for this fortunate
coincidence is explained in \cite{Robledo2025} where it is also warned that
the formulas in \cite{Bertsch2012} for overlaps of general operators are
not valid in the present case.
On the other hand, the HF formalism used in Ref
\cite{PhysRevLett.105.192701} is correct as the finitude of the
basis is automatically accounted for by the determinant of the overlap
matrix between the different orbitals involved \cite{Loewdin1955a}.
To our knowledge the formalism for finite basis not closed under rotations
\cite{Robledo1994,Rob22} has only been applied to the calculation
of angular momentum projected potential energy surfaces for fission
in \cite{Marevic2020}, the angular momentum distribution in fragments
\cite{Marevic2021}, and the evaluation of overlaps in fission when
the oscillator bases have different values of the oscillator lengths
\cite{Robledo2022}. The purpose of this paper is to develop the
formalism of \cite{Robledo1994} in the present context of particle
number restoration with BCS or HFB wave functions to provide the correct
expressions for the calculation of particle number distribution probabilities.
It is also proven how the present formalism provides the correct HF
limit whereas the incorrect one of Refs \cite{PhysRevC.100.024612,PhysRevC.100.034612,PhysRevC.101.014310,PhysRevC.103.054602,PhysRevC.104.054601,PhysRevLett.126.142502}
does not. The differences between the two methods are illustrated
in a simple one-dimensional example in the appendix.

Recently, angular momentum projection in a domain that is also widely
discussed in the literature \cite{PhysRevC.104.054601,PhysRevLett.126.142502,Marevic2021}
to analyze the distribution of angular momentum in the fragments.
The interest being triggered by experimental results obtained recently
\cite{Wilson2021}. In this case, one also needs the formalism presented
here (used in \cite{Marevic2021} but whitout a detailed explanation)
that is expected to substantially differ from the one in \cite{PhysRevC.104.054601,PhysRevLett.126.142502}. 

\section{Particle number projection on a domain}

To illustrate how projection on a domain $\mathcal{D}$ works in the
case of using HFB wave functions, the case of particle number will
be discussed in the following. To characterize the domain it is convenient
to introduce the idempotent operator 
\[
\hat{\varTheta}_{\mathcal{D}}(\vec{r})=\begin{cases}
\mathbb{I} & \vec{r}\subset\mathcal{D}\\
0 & \mathrm{otherwise}
\end{cases}
\]
which is the identity operator if $\vec{r}\subset\mathcal{D}$ and
zero otherwise. The particle number operator in the domain is defined
as the product, conveniently symmetrized, of $\hat{N}$ times $\hat{\Theta}_{\mathcal{D}}$
\[
\hat{N}_{\mathcal{D}}=\hat{\varTheta}_{\mathcal{D}}^{+}(\vec{r})\hat{N}\hat{\varTheta}_{\mathcal{D}}(\vec{r})
\]
 The operator is used to define the projector on particle number in
the domain $\mathcal{D}$
\begin{equation}
P_{\mathcal{D}}^{N}=\frac{1}{2\pi}\int_{0}^{2\pi}d\varphi\,e^{i\varphi(\hat{N}_{\mathcal{D}}-N)}\label{eq:Proj}
\end{equation}
The treatment becomes more transparent if operators are expressed
in terms of the creation and annihilation operators written in coordinate
representation
\begin{align*}
\hat{\psi}^{+}(\vec{r}) & =\sum_{k}\phi_{k}^{*}(\vec{r})c_{k}^{+}\\
\hat{\psi}(\vec{r}) & =\sum_{k}\phi_{k}(\vec{r})c_{k}
\end{align*}
where $\phi_{k}(\vec{r})$ is the wave function associated to the
state $|k\rangle=c_{k}^{+}|-\rangle$. The operator $\hat{N}_{\mathcal{D}}$
is given by
\[
\hat{N}_{\mathcal{D}}=\int d^{3}\vec{r}\hat{\psi}^{+}(\vec{r})\hat{\psi}(\vec{r})\varTheta_{\mathcal{D}}(\vec{r})
\]
where now $\varTheta_{\mathcal{D}}(\vec{r})$ is the Heaviside or
step function in the domain $\mathcal{D}$. It is also easy to obtain
\[
e^{i\varphi\hat{N}_{\mathcal{D}}}=\hat{\varTheta}_{\bar{\mathcal{D}}}(\vec{r})+e^{i\varphi}\hat{\varTheta}_{\mathcal{D}}(\vec{r})=\mathbb{I}+(e^{i\varphi}-1)\hat{\varTheta}_{\mathcal{D}}(\vec{r})
\]
where $\hat{\varTheta}_{\bar{\mathcal{D}}}(\vec{r})$ is the step
function in the complementary domain $\bar{\mathcal{D}}$ and it is
such that $\hat{\varTheta}_{\bar{\mathcal{D}}}(\vec{r})+\hat{\varTheta}_{\mathcal{D}}(\vec{r})=\mathbb{I}$.
The result of acting with the above operator on a wave function of
the basis $\phi_{k}(\vec{r})$ is to multiply it by $1+(e^{i\varphi}-1)\varTheta_{\mathcal{D}}(\vec{r})$
to give another wave function $\tilde{\phi}_{k}(\vec{r})=\left(1+(e^{i\varphi}-1)\varTheta_{\mathcal{D}}(\vec{r})\right)\phi_{k}(\vec{r}).$
These considerations use operators that are implicitly defined in
the whole Hilbert space and therefore they are implicitly associated
to a complete basis. When the basis used in the calculation is not
complete, as it is usually the case in practical calculations where
the number of basis states is finite, the new set of wave functions
$\tilde{\phi}_{k}(\vec{r})$ cannot be expressed in general as a linear
combination of the basis's elements. This is a fundamental remark
as the operators defined in the subspace spanned by the finite basis
do not share in general the same properties of the operators defined
in the whole space. In those cases the overlap between HFB states
and the associated generalized Wick's theorem have to be formulated
using the method discussed in Ref \cite{Rob22,Robledo1994} where
a formula is given for the overlap between HFB states which are expressed
in different bases not connected by a unitary transformation. This
situation might seems paradoxical because $\hat{N}_{\mathcal{D}}$
is hermitian and its exponential (multiplied by $i$ ) must be a unitary
operator. However, this line of reasoning is only correct when considering
the complete Hilbert space because the operator $\hat{N}_{\mathcal{D}}$
has non-zero matrix elements connecting every corner of the Hilbert
space - see Appendix A for an example. 

\subsection{Projected overlap}

Given a general HFB wave function $|\Psi\rangle$ the goal is to determine
$|\Psi_{\mathcal{D}}^{N}\rangle=\hat{P}_{\mathcal{D}}^{N}|\Psi\rangle$
in order to compute the particle number distribution probability
\begin{equation}
a(N)=\langle\Psi|\Psi_{\mathcal{D}}^{N}\rangle=\frac{1}{2\pi}\int_{0}^{2\pi}d\varphi\,e^{-i\varphi N}\langle\Psi|e^{i\varphi\hat{N}_{\mathcal{D}}}|\Psi\rangle.\label{eq:an}
\end{equation}
of the subsystem enclosed in $\mathcal{D}$. According to Refs \cite{Robledo1994,Rob22}
the more general overlap $\langle\Psi_{0}|e^{i\varphi\hat{N}_{\mathcal{D}}}|\Psi_{1}\rangle$
is given by
\begin{equation}
\langle\Psi_{0}|e^{i\varphi\hat{N}_{\mathcal{D}}}|\Psi_{1}\rangle=\sqrt{\det A\det\mathcal{R}}\label{eq:over}
\end{equation}
with
\begin{equation}
A=\bar{U}_{0}^{T}\left(\mathcal{R}^{T}\right)^{-1}\bar{U}_{1}^{*}+\bar{V}_{0}^{T}\mathcal{R}\bar{V}_{1}^{*}\label{eq:A}
\end{equation}
and 
\begin{equation}
\mathcal{R}_{nm}=\langle\phi{}_{n}|e^{i\varphi\hat{N}_{\mathcal{D}}}|\phi{}_{m}\rangle=\delta_{kl}+(e^{i\varphi}-1)\langle\phi{}_{n}|\hat{\varTheta}_{\mathcal{D}}|\phi{}_{m}\rangle.\label{eq:R}
\end{equation}
 In the above expressions $\bar{U}_{i}$ and $\bar{V}_{i}$ ($i=0,1)$
are the Bogoliubov amplitudes of each of the HFB wave functions (the
bar over the symbols is kept for consistence with the notation of
\cite{Rob22}) and $|\phi_{n}\rangle$ is a generic state of the
basis $B$ of dimension $N$, which is assumed to be the same for
both $|\Psi_{0}\rangle$ and $|\Psi_{1}\rangle$. The matrix $\left(\varTheta_{\mathcal{D}}\right)_{nm}=\langle\phi{}_{n}|\hat{\varTheta}_{\mathcal{D}}|\phi{}_{m}\rangle$
is hermitian and therefore can be diagonalized by a unitary transformation
$T$ to give real eigenvalues $\theta_{l}$. The operator $\hat{\varTheta}_{\mathcal{D}}$
is idempotent when defined in the whole Hilbert space and therefore
it should have eigenvalues $\theta_{l}$ equal to zero or one. However
this property does not apply in the case of a finite basis as the
idempotent property requires a complete basis - see Appendix A. The
overlap matrix $\mathcal{R}$ is diagonal also in the basis given
by the $T$ transformation $\mathcal{R}=TrT^{+}$ with eigenvalues
\begin{equation}
r_{l}(\varphi)=1+(e^{i\varphi}-1)\theta_{l}.\label{eq:rdiag}
\end{equation}
From here it is straightforward to write $\left(\mathcal{R}^{T}\right)^{-1}=T^{*}r^{-1}T^{T}$.
Please note that $r_{l}$ is not a phase if $\theta_{l}$ is different
from zero or one - see Appendix A. This is manifestation of the use
of finite basis not closed under the symmetry transformation. It is
worth to emphasize at this point that the matrix $A$ of Eq \ref{eq:A}
is not the one implicitly used in Refs \cite{PhysRevC.100.024612,PhysRevC.100.034612,PhysRevC.101.014310,PhysRevC.103.054602,PhysRevC.104.054601,PhysRevLett.126.142502}
(the $\left(\mathcal{R}^{T}\right)^{-1}$ here is replaced by $\mathcal{R}^{*}$
there) and therefore the conclusions discussed in those papers, obtained
in calculations with a finite basis, and in the case of non-zero pairing
states are not correct. Please note that in the ``standard'' formulation
for the particle number operator the overlap is given by 
\[
\langle\Psi|e^{i\varphi\hat{N}}|\Psi\rangle=e^{i\varphi\mathrm{Tr}[N]/2}\sqrt{\det A}
\]
with $A=e^{-i\varphi}\bar{U}^{T}\bar{U}^{*}+e^{i\varphi}\bar{V}^{T}\bar{V}^{*}$.
It is customary to factor the $e^{-i\varphi}$ in $A$ to cancel out
the phase in the overlap to write $\langle\Psi|e^{i\varphi\hat{N}}|\Psi\rangle=\sqrt{\det\bar{A}}$
with $\bar{A}=\bar{U}^{T}\bar{U}^{*}+e^{2i\varphi}\bar{V}^{T}\bar{V}^{*}$
. This is the strategy in \cite{PhysRevC.100.024612,PhysRevC.100.034612,PhysRevC.101.014310,PhysRevC.103.054602,PhysRevC.104.054601,PhysRevLett.126.142502}
that is clearly not justified in the present case of particle number
on a domain. On the other hand, the formulas and results using Slater
determinants \cite{PhysRevLett.105.192701}, relying on the formulas
derived in Ref \cite{Loewdin1955}, are correct. In the next subsection
I will prove that the present formalism reduces in the zero-pairing
limit to the result of Simenel and it will also be proven that this
is not the case for the formulas in \cite{PhysRevC.100.024612,PhysRevC.100.034612,PhysRevC.101.014310,PhysRevC.103.054602,PhysRevC.104.054601,PhysRevLett.126.142502}. 

\subsection{The zero-pairing limit}

As the formalism used in the HF case \cite{PhysRevLett.105.192701},
based on the formulas of \cite{Loewdin1955}, is correct, it is important
to check that the zero-pairing limit of the formalism presented above
recovers Simenel's results. In \cite{Robledo1994}, the zero-pairing
limit of Eq \ref{eq:over} was discussed and by using a simple argument
it was proven that the Löwdin formula \cite{Loewdin1955} is recovered
in this limit. Here we consider another derivation which explicitly
shows the different cancellations required to go from Eq \ref{eq:over}
to Löwdin's formula used in \cite{PhysRevLett.105.192701} and also
allows to see how the incorrect approaches of Refs \cite{PhysRevC.100.024612,PhysRevC.100.034612,PhysRevC.101.014310,PhysRevC.103.054602,PhysRevC.104.054601,PhysRevLett.126.142502}
do not reproduce the correct formula. The limit is easily obtained
by using Bloch-Messiah theorem \cite{Bloch1962,Zumino1962} providing
the canonical form of the Bogoliubov amplitudes
\begin{align}
\bar{U}_{i} & =D_{i}\bar{u}_{i}C_{i}\label{eq:BM}\\
\bar{V}_{i} & =D_{i}^{*}\bar{v}_{i}C_{i}\nonumber 
\end{align}
in terms of the unitary transformation to the canonical basis $D_{i}$,
the unitary transformation among quasiparticles $C_{i}$ and the block
diagonal matrices $\bar{u}_{i}$ and $\bar{v}_{i}$. The matrix $A$
of Eq \ref{eq:A} becomes
\[
A=C_{0}^{T}\left(\bar{u}_{0}^{T}(\bar{\mathcal{R}}^{T})^{-1}\bar{u}_{1}^{*}+\bar{v}_{0}^{T}\bar{\mathcal{R}}\bar{v}_{1}^{*}\right)C_{1}^{*}=C_{0}^{T}\bar{A}C_{1}^{*}
\]
where $\bar{\mathcal{R}}=D_{0}^{+}\mathcal{R}D_{1}$ and the matrix
$\bar{A}$ is the BCS version of Eq \ref{eq:A}. Up to an irrelevant
phase that becomes 1 when $|\Psi_{0}\rangle=|\Psi_{1}\rangle$\footnote{Any overlap involving non equal $|\Psi_{0}\rangle$ and $|\Psi_{1}\rangle$
is always defined up to arbitrary phases multiplying $|\Psi_{0}\rangle$
and $|\Psi_{1}\rangle$ } and that can be chosen by fixing the phase of $\langle\Psi_{0}|\Psi_{1}\rangle$,
one has $\det A=\det\bar{A}$ and $\det\mathcal{R}=\det\bar{\mathcal{R}}$
and therefore the formula for the overlap becomes $\langle\Psi_{0}|e^{i\varphi\hat{N}_{\mathcal{D}}}|\Psi_{1}\rangle=\sqrt{\det\bar{A}\det\bar{\mathcal{R}}}$.
The next step is to consider the expression of $\bar{\mathcal{R}}$
in block form
\begin{equation}
\bar{\mathcal{R}}=\left(\begin{array}{cc}
\bar{\mathcal{R}}_{11} & \bar{\mathcal{R}}_{12}\\
\bar{\mathcal{R}}_{21} & \bar{\mathcal{R}}_{22}
\end{array}\right)\label{eq:Rblock}
\end{equation}
where the matrix $\bar{\mathcal{R}}_{11}$ is the $A\times A$ matrix
(here $A$ is the number of particles) of the overlaps between the
occupied $A$ orbitals in the canonical basis of $|\Psi_{0}\rangle$
and those of $e^{i\varphi\hat{N}_{\mathcal{D}}}|\Psi_{1}\rangle$.
Next, one needs the formulas for the determinant 
\[
\det\bar{\mathcal{R}}=\det\bar{\mathcal{R}}_{11}\det\left[\bar{\mathcal{R}}_{22}-\bar{\mathcal{R}}_{21}\bar{\mathcal{R}}_{11}^{-1}\bar{\mathcal{R}}_{12}\right]
\]
and inverse 
\[
\left(\bar{\mathcal{R}}^{T}\right)^{-1}=\left(\begin{array}{cc}
\bullet & \bullet\\
\bullet & \left[\bar{\mathcal{R}}_{22}^{T}-\bar{\mathcal{R}}_{12}^{T}\left(\bar{\mathcal{R}}_{11}^{T}\right)^{-1}\bar{\mathcal{R}}_{21}^{T}\right]^{-1}
\end{array}\right)
\]
of the block matrix of Eq (\ref{eq:Rblock}). In the zero-pairing
limit the $\bar{u}_{0}=\bar{u}_{1}$ matrices are diagonal with 0
in the first $A$ diagonal elements and 1 in the remaining ones, and
the matrices $\bar{v}_{0}=\bar{v}_{1}$ are block diagonal with the
first $A$ elements consisting of $2\times2$ matrices $\left(\begin{array}{cc}
0 & 1\\
-1 & 0
\end{array}\right)$ or $1$ and the rest of the diagonal elements 0. Taking this form
into account one obtains
\[
\bar{A}=\left(\begin{array}{cc}
I_{v}^{T}\bar{\mathcal{R}}_{11}I_{v} & 0\\
0 & \left[\bar{\mathcal{R}}_{22}^{T}-\bar{\mathcal{R}}_{12}^{T}\left(\bar{\mathcal{R}}_{11}^{T}\right)^{-1}\bar{\mathcal{R}}_{21}^{T}\right]^{-1}
\end{array}\right)
\]
where $I_{v}$ is the $A\times A$ diagonal matrix with the first
$A$ elements consisting of $2\times2$ matrices $\left(\begin{array}{cc}
0 & 1\\
-1 & 0
\end{array}\right)$ or $1$. As $\det I_{v}=1$ one finally gets
\[
\det\bar{A}\det\bar{\mathcal{R}}=\left(\det\bar{\mathcal{R}}_{11}\right)^{2}
\]
which gives, in the zero pairing limit and up to a phase
\[
\langle\Psi_{0}|e^{i\varphi\hat{N}_{\mathcal{D}}}|\Psi_{1}\rangle=\det\bar{\mathcal{R}}_{11}.
\]
In the $|\Psi_{0}\rangle=|\Psi_{1}\rangle$ case where the phase is
1 one recovers Löwdin formula as used in \cite{PhysRevLett.105.192701}.
Please note that if we were using the incorrect formulas of \cite{PhysRevC.100.024612,PhysRevC.100.034612,PhysRevC.101.014310,PhysRevC.103.054602,PhysRevC.104.054601,PhysRevLett.126.142502}
we should replace $(\bar{\mathcal{R}}^{T})^{-1}$ by $\bar{\mathcal{R}}^{*}$
and therefore in that case $\det\bar{A}\det\bar{\mathcal{R}}=\left(\det\bar{\mathcal{R}}_{11}\right)^{2}\det\bar{\mathcal{R}}_{22}^{*}/\det\left[\bar{\mathcal{R}}_{22}^{T}-\bar{\mathcal{R}}_{12}^{T}\left(\bar{\mathcal{R}}_{11}^{T}\right)^{-1}\bar{\mathcal{R}}_{21}^{T}\right]$
which is not the correct limit. 

\subsection{The Pfaffian formula}

In the previous formalism the overlap is given by a square root and
therefore its sign is not defined in general \cite{Robledo2009}.
This can become a real problem when the overlap depends on a continuous
variable used as an integration variable. To circumvent the problem
it was proposed to use a new formula based on techniques using fermion
coherent states \cite{Robledo2009,Robledo2011,Bertsch2012}. The
formula, commonly denoted as ``pfaffian formula'' is given in terms
of the Pfaffian of a skew symmetric matrix and it is formulated assuming
the two HFB states are expressed in terms of the same basis (or equivalent
bases up to a unitary transformation). When this is not the case,
an extension of the pfaffian formalism given in Ref \cite{Robledo2011}
solves the problem. In the following we will adapt the results of
Ref \cite{Robledo2011} to our present situation. As discussed in
\cite{Robledo2011} the relevant quantity is the extended overlap
matrix
\[
\mathcal{N}=\left(\begin{array}{cc}
\mathbb{I} & \mathcal{R}\\
\mathcal{R}^{+} & \mathbb{I}
\end{array}\right)
\]
defined in the vector space union of the two bases $B=\{|\phi_{n}\rangle,n=0,\ldots,N-1\}$
and $\tilde{B}=\{|\tilde{\phi}_{n}\rangle=e^{i\varphi\hat{N}_{\mathcal{D}}}|\phi_{n}\rangle,n=0,\ldots,N-1\}$.
The matrix $\mathcal{R}$ is the one of Eq \ref{eq:R} with eigenvalues
given in Eq \ref{eq:rdiag} in terms of the real eigenvalues $\theta_{l}$
of the matrix $\left(\varTheta_{\mathcal{D}}\right)_{nm}=\langle\phi{}_{n}|\hat{\varTheta}_{\mathcal{D}}|\phi{}_{m}\rangle$.
Adapting the developments in Ref \cite{Robledo2011} to the present
set up one obtains
\begin{equation}
\langle\Psi_{0}|e^{i\varphi\hat{N}_{\mathcal{D}}}|\Psi_{1}\rangle=(-1)^{N}\textrm{pf}\left(\begin{array}{cc}
\tilde{N}_{D}^{(1)} & -\mathbb{I}\\
\mathbb{I} & -\tilde{N}_{D}^{(0)\,*}
\end{array}\right)\label{eq:Overlap_F}
\end{equation}
where
\begin{equation}
\tilde{N}_{D}^{(0)}=\frac{1}{2}\left(\begin{array}{cc}
E_{+} & 0\\
0 & E_{-}
\end{array}\right)\left(\begin{array}{cc}
M^{(0)} & -M^{(0)}\\
-M^{(0)} & M^{(0)}
\end{array}\right)\left(\begin{array}{cc}
E_{+}^{T} & 0\\
0 & E_{-}^{T}
\end{array}\right)\label{eq:Ntilda0DFinal}
\end{equation}
and
\begin{equation}
\tilde{N}_{D}^{(1)}=\frac{1}{2}\left(\begin{array}{cc}
F_{+} & 0\\
0 & F_{-}
\end{array}\right)\left(\begin{array}{cc}
M^{(1)} & M^{(1)}\\
M^{(1)} & M^{(1)}
\end{array}\right)\left(\begin{array}{cc}
F_{+}^{T} & 0\\
0 & F_{-}^{T}
\end{array}\right).\label{eq:Ntilda1DFinal}
\end{equation}
In the above expressions, $E_{\pm}=n_{\pm}^{1/2}T^{+}$ and $F_{\pm}=n_{\pm}^{1/2}f_{r}T^{+}$
with $T$ the matrix diagonalizing $\left(\varTheta_{\mathcal{D}}\right)_{nm}$.
The diagonal matrices $n_{\pm}$ contain the diagonal elements $n_{\pm}=1\pm|r|$
where the complex quantities $r_{l}$ are defined in Eq \ref{eq:rdiag}
and $f_{r}$ is the phase of the complex number $r$. Finally, the
matrices $M^{(i)}$ are the matrices of the Thouless transformation
defining $|\Psi_{i}\rangle$ (see \cite{Robledo2011} for detailed
definitions). In the limiting case where the matrix $\left(\varTheta_{\mathcal{D}}\right)_{nm}$
is the identity, its eigenvalues are $1,$ the $r_{l}$ are all the
same $r_{l}=e^{i\varphi}=f_{r}$, $|r|=1$ and $n_{+}=2$, $n_{-}=0$.
With these values one recovers the results of \cite{Robledo2009,Bertsch2012}.

\subsection{The integration limit in the projector}

The particle number projector is defined in general in Eq (\ref{eq:Proj})
with the $2\pi$ upper limit in the integral. It is often the case
that the integration limit is reduced to the interval $[0,\pi]$ by
assuming that $e^{i\pi(\hat{N}_{\mathcal{D}}-N)}$ is the identity
when acting on a HFB state due to ``number parity'' super-selection
rule. This statement is strictly true when the domain $\mathcal{D}$
is the whole space but it is not correct for a finite domain. For
this property to be satisfied it is required the eigenvalues $r_{l}$
of $\mathcal{R}$ (Eq (\ref{eq:rdiag})) to be $-1$ when $\varphi=\pi$
which is only possible if $\theta_{l}=1$. Therefore, it is mandatory
to extend the integral in the projector to $2\pi$ as in this case
the correct value of $r_{l}=1$ is recovered irrespective of the value
of $\theta_{l}$. . On the other hand, owing to the property $r_{l}(2\pi-\varphi)=r_{l}^{*}(\varphi)$
one can reduce the integration interval but at the price of replacing
the integrand by twice its real part. .

\subsection{Reinterpretation of the formalism}

In order to get more insight into the formalism it is convenient to
introduce some ideas of condense matter physics and quantum information
\cite{Klich2006}, first introduced in the context of fission in
\cite{Qiang2025}. Let us define the non-orthogonal states $\tilde{\phi}_{k}^{\mathcal{D}}(\vec{r})=\Theta_{\mathcal{D}}(\vec{r})\phi_{k}(\vec{r})$
and their complementary $\tilde{\phi}_{k}^{\bar{\mathcal{D}}}(\vec{r})=\Theta_{\bar{\mathcal{D}}}(\vec{r})\phi_{k}(\vec{r})$
such that $\phi_{k}=\tilde{\phi}_{k}^{\mathcal{D}}+\tilde{\phi}_{k}^{\bar{\mathcal{D}}}$.
One can orthogonalize those states by diagonalizing the hermitian
matrix $\left(\varTheta_{\mathcal{D}}\right)_{nm}$ with eigenvalues
$\theta_{l}$ to obtain the set of orthogonal states $\phi_{k}^{R}(\vec{r})=1/\sqrt{\theta_{k}}\sum_{l}T_{lk}\tilde{\phi}_{l}^{\mathcal{D}}(\vec{r})$
and $\phi_{k}^{L}(\vec{r})=1/\sqrt{1-\theta_{k}}\sum_{l}T_{lk}\tilde{\phi}_{l}^{\bar{\mathcal{D}}}(\vec{r})$
which are also orthogonal for every $k$ and $l$, i.e. $\langle\phi_{k}^{R}|\phi_{l}^{L}\rangle=0$.
The set of orthogonal states $|\phi_{k}^{R}\rangle=R_{k}^{+}|\rangle$
are non-zero in the interior of the domain $\mathcal{D}$ whereas
$|\phi_{k}^{L}\rangle=L_{k}^{+}|\rangle$ have the same property but
in the complementary $\bar{\mathcal{D}}$. Please note zero (one)
eigenvalues have to be excluded in the set $\phi_{k}^{R}$ ($\phi_{k}^{L}$).
With these definitions the basis states $|\phi_{k}\rangle=c_{k}^{+}|\rangle$
can be expanded as a linear combination 
\begin{equation}
c_{l}^{+}=\sum_{k}T_{lk}^{*}\left[\sqrt{\theta_{k}}R_{k}^{+}+\sqrt{1-\theta_{k}}L_{k}^{+}\right]\label{eq:RL}
\end{equation}
of states on the domain and outside it. This definition gives a new
perspective to the values of $\theta_{k}$ (see appendix below). The
$\hat{N}_{\mathcal{D}}$ operator can be written as $\hat{N}_{\mathcal{D}}=\sum_{k}R_{k}^{+}R_{k}$
what leads to $e^{i\varphi\hat{N}_{\mathcal{D}}}c_{l}^{+}e^{-i\varphi\hat{N}_{\mathcal{D}}}=\sum_{k}T_{lk}^{*}[\sqrt{\theta_{k}}e^{i\varphi}R_{k}^{+}+\sqrt{1-\theta_{k}}L_{k}^{+}]$.
This result shows that the ``number parity'' associated to the $c_{l}^{+}$
operators is not respected by the ``number parity'' of the $R_{k}^{+}$
implying, for instance, that a Slater determinant with an even number
of particles has a non zero amplitude of having an odd number of particles
in the domain $\mathcal{D}$. In practical terms, the probability
amplitude $a(N)$ of a system with an even number of particles can
also be different from zero for odd values of N. From a quantum information
perspective one can say that the original wave function is entangled
with the one defined in the domain \cite{Klich2006}. 

\section{Application in a simple case}

In order to appreciate the differences between the results obtained
with the correct and the incorrect formalism as well as to study the
entanglement in the system, I have carried calculations in a simple
1D model where the Hilbert space basis contains $N_{B}=40$ elements
corresponding to the spin $1/2,$ 1D Harmonic Oscillator wave functions
with oscillator length $b=1$ and with quantum numbers $n=0,\ldots,N_{B}/2-1$.
The spin degree of freedom is explicitly considered giving a degeneracy
factor of 2. For the pairing part, I have chosen a set of occupancies
$v_{n}^{2}=1/(1+\exp((n-n_{0})/\sigma))$ where $n_{0}$ and $\sigma$
fix the shape of the orbital's occupancy. The particle number probability
distribution $a(N)$ has been computed with both the correct and the
incorrect formulas in various situations including different domains
$\theta(x-x_{0})$ with various values of $x_{0}$. The zero-pairing
limit is also considered. The results, shown in appendix A clearly
show how different the $a(N)$ are except in those trivial cases where
they should coincide (large negative values of $x_{0}$ to include
in an effective way the whole Hilbert space). Finally, a more realistic
1D model is devised by using a Bogoliubov transformation tailored
to produce a matter density with two fragments. Also in this case
the $a(N)$ differ pointing again to the necessity of considering
the formalism of \cite{Robledo1994,Rob22} when dealing with finite
size basis. 

\begin{figure}
\includegraphics[width=0.47\columnwidth]{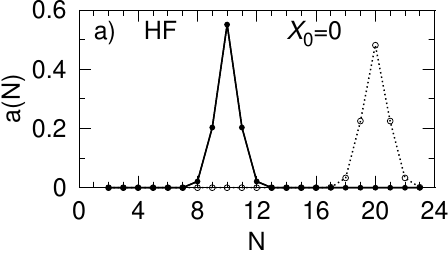}\includegraphics[width=0.47\columnwidth]{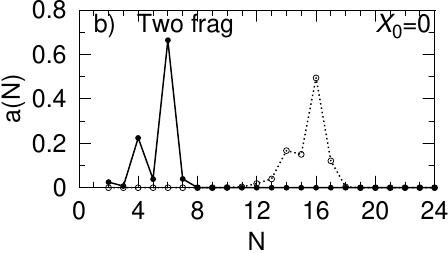}\caption{Particle number distribution $a(N)$ as a function of N in a) the
zero-pairing limit and b) the two-fragment solution. In both panels
the domain correspond to $x_{0}=0$ (i.e. positive values of $x$)
and the results corresponding to the correct formulation are given
by full line and bullets whereas the ones of the incorrect formulation
are given by circles and dotted lines. In panel a) the result obtained
with Löwdin formula is also presented but it coincides with the correct
one. In the two-fragment case the average number of particles in the
fragment to the right is 5.4. \protect\label{fig:Fig2}}
\end{figure}
Although the model is discussed in detail in the appendix we show
in Fig \ref{fig:Fig2} two examples of $a(N)$ distributions obtained
in the zero-pairing limit and the two-fragment case. In both cases
the correct distribution substantially differ from the incorrect one.
The correct one is always peaked at the value of the average number
of particles in the fragment. In the no-pairing case it is remarkable
that $a(N)$ significantly differs from zero for odd values of N as
a consequence of the entanglement with the fragment of the total wave
function. On the other hand, in the two-fragment case the correct
$a(N)$ is very small for odd values of N indicating a very low degree
of entanglement associated to the clean definition of the fragment
in this situation.

\section{Summary and conclusions }

In this paper we discuss how to compute properly the overlaps required
in the projection on particle number on a domain $\mathcal{D}$ when
dealing with HFB wave functions. It is shown that the formulas used
in a series of recent papers are incorrect and the conclusions obtained
in those papers dubious. The reason for the pitfall is not taking
into account that the action of the symmetry operator on the finite
basis leads to another basis which is not unitarily equivalent. The
situation is exemplified in a simple case demonstrating the non-equivalence
of some results when dealing with finite bases. Additionally, is is
shown that the non-zero pairing limit of the present approach recovers
Löwdin formula, contrary to previous formulations by other authors.
The results discussed here call for a revision of a similar formalism
for angular momentum projection on a domain. 

\section*{Acknowledgements}

I would like to thank S.A.Giuliani for a carefull reading of the manuscript
and several suggestions. This work has been supported by Spanish Agencia
Estatal de Investigacion (AEI) of the Ministry of Science and Innovation
under Grant No. PID2021-127890NB-I00.

\appendix

\section{Finite basis: 1D example}

\begin{figure}
\includegraphics[width=0.95\columnwidth]{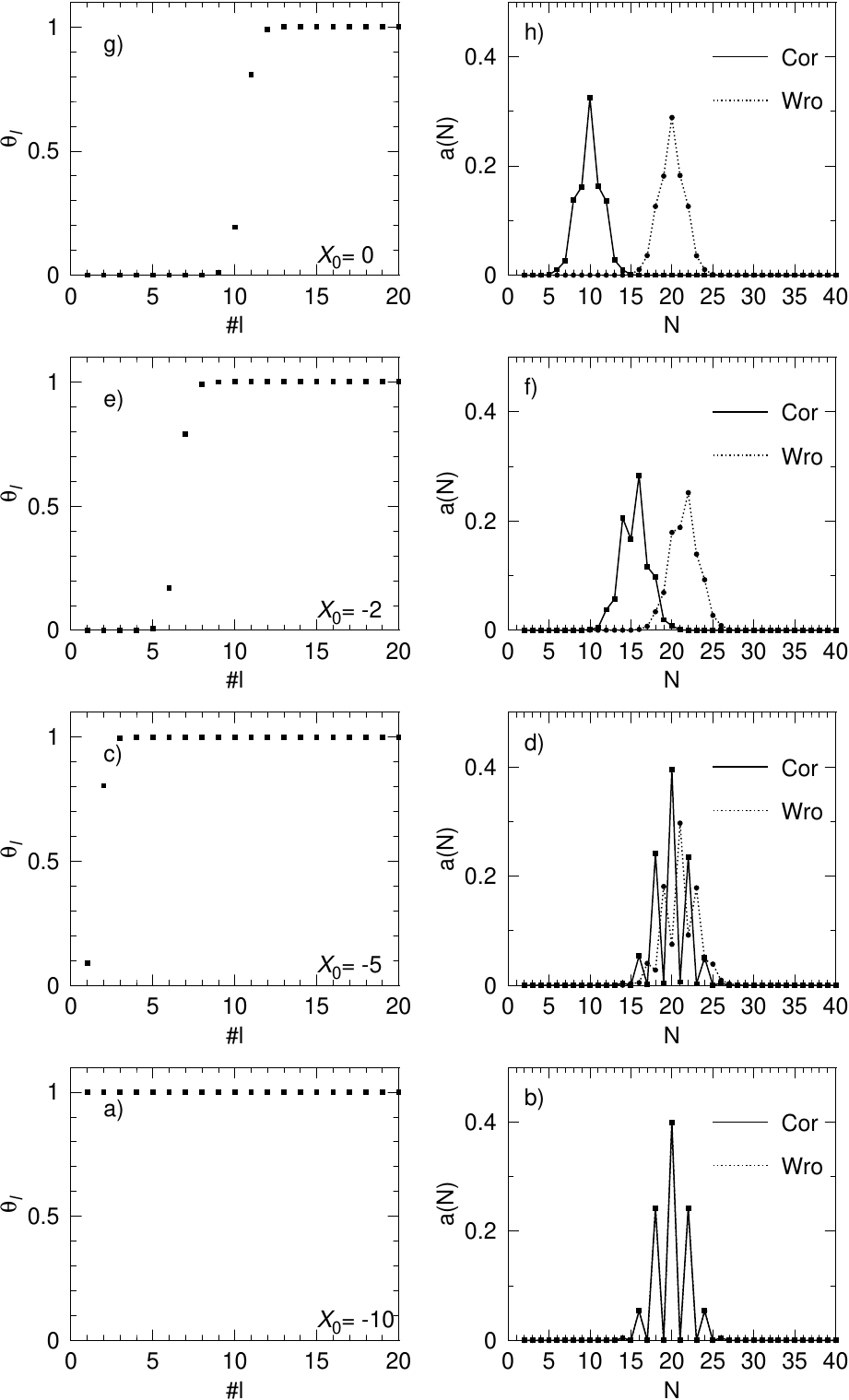}\caption{Left panels, distribution of eigenvalues $\theta_{l}$ for different
domains $\theta(x-x_{0})$ with $x_{0}=$-10,- 5, -2 and 0. Right
panels, the corresponding probability of finding a wave function with
particle number $N$ in the intrinsic wave function. The legend ``Corr''
corresponds to the results obtained with the correct formulas disucssed
in this paper whereas ``Wro'' stands for the incorrect results obtained
with the approaches discussed in the main text\protect\label{fig:Fig1}}
\end{figure}

To better visualize the problematic associated to the use of finite
basis one can consider a one-dimensional Harmonic Oscillator (HO)
basis $B_{N}=\{\phi_{n}(x),n=0,\ldots,N-1$\} and a domain $\mathcal{D}$
corresponding to positive values of $x$. The action of $e^{i\varphi\hat{N}_{\mathcal{D}}}$
on the basis' states leads to non-continuous, non-diferentiable wave
function at $x=0$ 
\[
\tilde{\phi}_{n}(x)=e^{i\varphi\hat{N}_{\mathcal{D}}}\phi_{n}(x)=\begin{cases}
\phi_{n}(x) & x<0\\
e^{i\varphi}\phi_{n}(x) & x\geq0
\end{cases}
\]
such that the scalar product with a general state $\phi_{m}(x)$ is
given by 
\[
\langle\phi{}_{n}|\tilde{\phi}{}_{m}\rangle=\delta_{nm}+\left(e^{i\varphi}-1\right)\theta_{nm}
\]
with 
\begin{align}
\theta_{nm} & =\int_{0}^{\infty}dx\phi_{n}^{*}(x)\phi_{m}(x)\nonumber \\
 & =\sqrt{\frac{2^{n+m}n!m!}{4\pi}}\sum_{pq}\frac{(-)^{p+q}\Gamma((n+m-2p-2q)/2)}{2^{2p+2q}p!q!(n-2p)!(m-2q)!}.\label{eq:theta}
\end{align}
This is a peculiar matrix with $\theta_{nn}=1/2$ and $\theta_{nm}=0$
if $n+m$ is even and $n\ne m.$ The matrix elements are non-zero
even when $n$ or $m$ (or both) are greater than $N$, i.e. $\tilde{\phi}_{n}(x)$
cannot be expressed as a linear combination of the states in basis
$B$. On the other hand, we will assume that such expansion is possible
in the case of using an infinite numerable basis. It is also worth
to remark that the new set $\tilde{\phi}_{n}(x)$ is orthonormal,
as the original one. This property is also preserved in the most general
case presented in the main text. The matrix $\theta_{nm}$ is hermitian
and therefore it can be diagonalized by a unitary transformation with
real eigenvalues $\theta_{l}$. In the simplest $N=2$ case one has
$\theta_{11}=\theta_{22}=1/2$ and $\theta_{12}=\theta_{21}=1/\sqrt{2\pi}$
and the eigenvalues are $\theta_{0}=0.1010$ and $\theta_{1}=0.8989$
, as can be easily checked. In the general case of a domain $\mathcal{D}$
consistent of $x\ge x_{0}$ the formula \ref{eq:theta} can be easily
generalized by considering incomplete gamma function. In order to
check the differences between our results and previous ones we have
performed calculations for simple BCS wave functions in the basis
$B_{N}$ with an occupation profile given by $v_{n}^{2}=1/(1+\exp(n+1-n_{0})/\sigma)$
with $N=20,$ $n_{0}=10.5$ and $\sigma=1$ giving an average number
of 20 particles. In Fig \ref{fig:Fig1} we plot the results corresponding
to $\theta_{l}$ and $a(N)$ for $x_{0}$ values of -10, -5, -2 and
0. For $x_{0}=-10$ the domain essentially covers all the $x$ values
where the spatial density (see below) is different from zero. As expected,
all the $\theta_{l}=1$ and our formulation and the incorrect one
given in the literature coincide with the present formulation. We
also observe that $a(N)=0$ for odd values of $N$ as expected. For
$x_{0}=-5$ a tiny part of the density is left off the domain (see
Fig \ref{fig:Fig3} below) and, as a consequence the two lowest $\theta_{l}$
values differ from the nominal values 0 or 1. We observe how the $a(N)$
probability strongly differs in the two cases with the incorrect one
peaked at N=22 and showing a completely different shape than the correct
one. It is worth to mention that the incorrect $a(N)$ gets its larger
values at odd values of N, contrary to the correct distribution where
the $a(N)$ for odd N are quite small. The total number of particles
in the domain, obtained by means of the formula for the particle number
sumrule $\sum_{N}a(N)\times N$ equals 19.98 (instead of 20) for the
correct formula and takes the value 21.08 for the incorrect one. For
$x_{0}=-2$ there are only two $\theta_{l}$ values different from
zero or one. Five $\theta_{l}$ values are zero and 13 equal 1 indicating
that the $R_{k}^{+}$ basis contains 15 vectors, whereas the $L_{k}^{+}$
one contains 7 basis states. The maximum of the two probability distribution
differ being the correct one centered at N=16 and the incorrect at
N=24. The number of particle sumrule is 15.47 and 21.51 for the correct
and incorrect results, respectively, wich is a manifiestation of strong
entanglement between the $x_{0}=-2$ domain and the rest of the space.
Finally, the $x_{0}=0$ corresponding to a domain including half of
the matter distribution have a similar behavior for $\theta_{l}$
and the probability distributions have similar shapes but their maximum
are clearly shifted being the correct $a(N)$ centered around the
correct value of 10 (remember that the average number of particles
in the whole system is 20) whereas in the incorrect case it is centered
around 20. These results clearly indicate the inconsitency of the
incorrect approach in estimating the probability distribution. In
the case the $a(N)$ for odd N are significantly different from zero
in the two cases.

It is also interesting to compare the no-pairing limit of the two
probabilities to the result obtained with Löwdin formula. The results
shown in Fig \ref{fig:Fig2}a in the main text indicate that the correct
formula gives exactly the same values than Löwdin formula as expected.
The $a(N)$ is peaked at N=10 as expected for $x_{0}=0$ and the particle
number sumrule is satisfied but with a large contribution of odd values
of N, indicating a large degree of entanglement in this case. The
incorrect formula gives a $a(N)$ peaked at $N=20$ which is clearly
not correct.

In order to better understand the situation we have performed an additional
calculation where we have used a discrete variable representation
(DVR) basis by diagonalizing the position operator in the harmonic
oscillator basis $B_{N}$ used before. The eigenvectors are labeled
by the values of the eigenvalues which are roughly uniformly distributed
in the $x$ axis. The occupancies in the new basis are taken as 0
for most of the states and $v_{k}^{2}=0.9$ for those with DVR mesh
points in between -4.6 and -2.25 (larger fragment) and in between
2.78 and 3.94 (smaller fragment). In this way we obtain the matter
distribution depecited in Fig \ref{fig:Fig3} which is shown along
with the matter density of the previouly discussed case. The total
number of particles is 14.4 whereas in the smaller fragment (to the
right) the number of particles is 5.4. The two fragments are well
separated and therefore entanglement effects are expected to be small. 

\begin{figure}

\includegraphics[width=0.95\columnwidth]{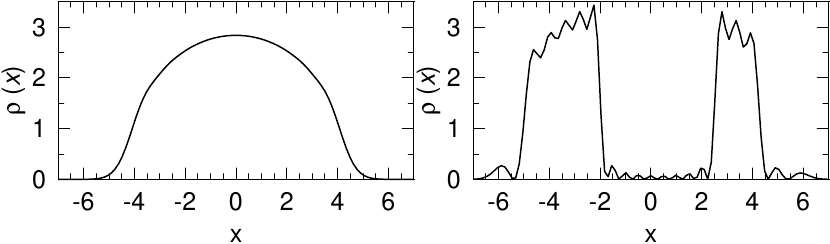}\caption{Matter density corresponding to the two cases studied here. The one
on the left is obtained by filling the lowest HO orbitals with a given
occupancy distribution whereas the one on the right is obtained by
using the DVR basis with ad-hoc occupancies used to obtain a density
corresponding to two fragments.\protect\label{fig:Fig3}}
\end{figure}

We finally provide the particle number probability distribution obtained
with the wave function corresponding to two separate fragments taking
as domain positive values of $x$, i.e. $x_{0}=0$. The particle number
probability distribution is depicted in Fig \ref{fig:Fig2}b for both
the correct formulation (full line) and the incorrect one (dotted
line). We observe how $a(N)$ peaks at $N=6$ in good agreement with
the average number of fragments in the right hand side fragment (5.4)
but this is not the case for the incorrect calculation. The number
of particle sumrule is in this case 5.4 which is the expected value
in absence of entanglement. This coincidence between the sum rule
and the actual average value is a consequence of having a well separated
fragment (see Fig \ref{fig:Fig3}) where the entanglement of the wave
function is not relevant. On the other hand, the incorrect formula
gives in this case a particle number sumrule of 15.42 which is clearly
incorrect.


\end{document}